\documentclass[aps,pra,reprint,superscriptaddress,amsmath,amssymb,showpacs]{revtex4-1}
\usepackage{bm}
\newtheorem{thm}{Theorem}

\begin{document}


\title{Optimal PPTES witnesses for states in $\mathbb C^n\otimes \mathbb C^n$}

\author{Kil-Chan Ha}
\affiliation{Faculty of Mathematics and Statistics, Sejong University, Seoul 143-747, Korea}
\date{\today}

\begin{abstract}
Recently, X. Qi and J. Hou [Phys. Rev. A {\bf 85}, 022334 (2012)] provided optimal entanglement witnesses without the spanning property. This witnesses are associated to indecomposable positive linear maps, but it is not checked whether partial transposes of these witnesses are also optimal. We show that partial transposes of these entanglement witnesses have spannning property, and these witnesses are indeed optimal PPTES witnesses (non-decomposable optimal entanglement witnesses).

\end{abstract}


\maketitle


\section{Introduction}

As it is well known, quantum entanglement is considered as a basic physical resource to realize various quantum information and quantum computation tasks. So, it is very important to distinguish entanglement from separable states. The most general approach for this purpose may be a criterion based on the notion of entanglement witness \cite{horo-1,terhal}. 
A Hermitian operator $W$ acting on a complex Hilbert space $\mathcal H\otimes \mathcal K$ is an entanglement witness (EW) if $W$ is not positive and $\text{Tr}(W\rho)\ge 0$ holds for all separable states $\rho$. 
This criterion  is equivalent to the duality theory \cite{eom-kye} between positivity of linear maps and
separability of block matrices, through the Jamio\l kowski-Choi
isomorphism \cite{choi75-10, jami}. That is, a self-adjoint block matrix $W\in M_n\otimes M_n$ is EW if and only if there exists a positive linear map that is not comletely positive  $\Phi: M_n\to M_n$ such that
\[
W=\frac 1n C_{\Phi}=\frac 1n\sum_{i,j=1}^n |i\rangle\langle j|\otimes \Phi(|i\rangle \langle j|),
\]
where $M_n$ denotes the $C^*$-algebra of all $n\times n$ matrices over the complex field $\mathbb C$ and the block matrix $C_{\Phi}$ is the Choi matrix of $\Phi$. We denote $W_{\Phi}=1/n\, C_{\Phi}$ for the entanglement witness associated to the positive map $\Phi$.

An entanglement witness which detects a maximal set of entanglement
is said to be optimal, as was introduced in \cite{lew00}. Lewenstein, Kraus, Cirac, and Horodecki \cite{lew00} showed that $W$ is an optimal entanglement witness if $W$ has spanning property, that is $\mathcal P_{W}=\{|\xi, \eta\rangle \in \mathbb C^n\otimes \mathbb C^m\,:\, \langle \xi,\eta |W|\xi,\eta\rangle=0\}$ spans the whole space $\mathbb C^n\otimes \mathbb C^m$. Note that the spanning property is not a necessary condition for optimality of EW since the Choi map \cite{choi75, choi-lam} gives rise to an optimal entanglement witness that have no spanning property. This gap can be explained in terms of facial structures of the convex cone $\mathbb P_1$ consisting of all positive linear maps between matrix algebras: An entanglement witness $W_{\Phi}$ is optimal if and only if the smallest face of $\mathbb P_1$ containing $\Phi$ has no completely positive linear map \cite{kye_dec_wit}, whereas 
an entanglement witness $W_{\Phi}$ has the spanning property if and only if the smallest exposed face of the cone $\mathbb P_1$ containing $\Phi$ has no completely positive linear map \cite{kye_ritsu}.
See also Ref.~\cite{hakye12}.

It is now well known that positive map $\Phi$ is indecomposable if and only if the associated EW $W_{\Phi}$ detects entangled states with positive partial transpose (PPTES). 
Recall that the EW associated to a positive linear map $\Phi$ is said to be a non-decomposable optimal entanglement witness(nd-OEW) in \cite{lew00} if it detects a maximal set of PPTES. It is clear \cite{lew00} that $W$ is nd-OEW if and only if both $W$ and $W^{\Gamma}$ are optimal, where $W^{\Gamma}$ denotes the partial tranpose of $W$.
We note that optimal entanglement witness associated to an indecomposable positive map may not be really nd-OEW in the sense of \cite{lew00}. In fact, it was shown \cite{hakye12} that there exist optimal entanglement witnesses which are associated to non-decomposable positive linear maps, but which are not nd-OEW in the sense \cite{lew00}. Thus, we suggested \cite{hakye12} the term {\sl PPTES witness} and {\sl optimal PPTES witness} in the places of \lq non-decomposable entanglement witness\rq\ and \lq non-decomposable optimal entanglement witness\rq in order to avoid possible confusion.

Recently,  X. Qi and J. Hou \cite{xia} provided a method of checking optimality of entanglement witnesses, and showed that indecomposable positive linear maps presented in \cite{xia11} give rise to indecomposable optimal entanglement witnesses which have no spanning property. But, it is not checked that these entanglement witnesses are indeed optimal PPTES witnesses (nd-OEW in the sense \cite{lew00}).

The purpose of this Brief Report is to show that the entanglement witnesses $W_{\Phi^{(n,k)}}$ provided by X. Qi and J. Hou are indeed optimal PPTES witnesses. First, we recall the indecomposable positive linear maps $\Phi^{(n,k)}:M_n\to M_n$ for each $k=1,2,\ldots,n-1$ defined by 
\begin{equation}\label{map}
\Phi^{(n,k)}([a_{ij}])=\text{diag}(b_1,b_2,\cdots,b_n)-[a_{ij}]
\end{equation}
for $[a_{ij}]\in M_n$, 
where $b_{i}=(n-1)a_{ii}+a_{k+i\,k+i}$ for each $i=1,2,\ldots,n$ $(n\ge 3)$, and subscript indices of matrix entries are understood by modulo $n$. We define matrices $A_{ij}^{(n,k)}\in M_n$ for each $k=1,2,\cdots, n-1$ by 
\[
A_{ij}^{(n,k)}=\begin{cases} 
(n-2)|i\rangle \langle i|+|n-k+i\rangle \langle n-k+i|&\ \text{if } i=j\\
-|i\rangle \langle j| & \ \text{if }i\neq j
\end{cases} 
\]
for $i,\,j=1,2,\cdots, n$, where we add modulo $n$. Then the associated EW $W_{\Phi^{(n,k)}}$ and its partial tranpose $W_{\Phi^{(n,k)}}^{\Gamma}$ are given by 
\[
\begin{aligned}
W_{\Phi^{(n,k)}}&=\sum_{i,j=1}^n |i\rangle \langle j|\otimes A_{ij}^{(n,k)},\\
W_{\Phi^{(n,k)}}^{\Gamma}&=\sum_{i,j=1}^n |i\rangle \langle j|\otimes A_{ji}^{(n,k)}.
\end{aligned}
\]

For any $n$-tuple $\theta=(\theta_1,\theta_2,\cdots, \theta_n)$ of real numbers $\theta_i$, we define a vector $|\xi_{\theta}\rangle \in \mathbb C^n$ by 
\begin{equation}\label{vec1}
|\xi_{\theta}\rangle=\sum_{k=1}^n e^{i\theta_k}|k\rangle.
\end{equation}
Then, it is well known \cite{xia} that product vectors $|\xi_{\theta}\otimes \xi_{\theta}^*\rangle $ belong to the set $\mathcal P_{W_{\Phi^{(n,k)}}}$ for any $n$-tuple $\theta$. Therefore, every product vector $|\xi_{\theta}\otimes \xi_{\theta}\rangle $ belongs to the set $\mathcal P_{W_{\Phi^{(n,k)}}^{\Gamma}}$, that is,
\[
\langle \xi_{\theta}\otimes \xi_{\theta}|W_{\Phi^{(n,k)}}^{\Gamma}|\xi_{\theta}\otimes \xi_{\theta}\rangle
=\langle \xi_{\theta}\otimes \xi_{\theta}^*|W_{\Phi^{(n,k)}}|\xi_{\theta}\otimes \xi_{\theta}^*\rangle=0
\]
for any $n$-tuple $\theta$.  We also define a set $\mathcal V_i^{(n,k)}$ of product vectors in $\mathbb C^n\otimes \mathbb C^n$ for each $k=1,2,\cdots, n-1$ and $i=1,2,\cdots,n$ as follows:
\begin{equation}\label{vec2}
\begin{aligned}
\mathcal V_i^{(n,k)}=\{ |i\rangle &\otimes |j\rangle:\, j\neq i \text{ and } \\ &j\neq n-k+i\ (\text{mod}\, n) \text{ for } 1\le j \le n\}.
\end{aligned}
\end{equation}
Note that each set $\mathcal V_i^{(n,k)}$ consists of $n-2$ product vectors.
It is easy to see that 
\[
\mathcal V_i^{(n,k)}\subset \mathcal P_{W_{\Phi^{(n,k)}}^{\Gamma}}
\]
for each $k=1,2,\cdots, n-1$ and $i=1,2,\cdots,n$. 

Now, we will show that vectors $|\xi_{\theta}\otimes \xi_{\theta}\rangle$ defined in Eq.~\eqref{vec1} and vectors in Eq.~\eqref{vec2} span the whole space $\mathbb C^n\otimes \mathbb C^n$ whenever $k\neq n/2$, and so each $W_{\Phi^{(n,k)}}^{\Gamma}$ is optimal whenever $k\neq n/2$. For the simplicity,  we identify $\mathbb C^n\otimes \mathbb C^n$ with $M_n$, that is, a product vector $\sum_{i=1}^n x_i|i\rangle \otimes \sum_{j=1}^n y_j|j\rangle$ is identified with
a matrix $\sum_{i,j=1}^n x_i y_j|i\rangle \langle j|$. 

First, we show that vectors $|\xi_{\theta}\otimes \xi_{\theta}\rangle$ in Eq.~\eqref{vec1} span all symmetric matrices in $M_n$ with this identification. For any $n$-tuple $\theta=(\theta_1,\theta_2,\cdots,\theta_n)$ and integer $\ell$, we denote $\theta^{(\ell)}=(\ell\, \theta_1,\ell\, \theta_2,\cdots,\ell\, \theta_n)$, and  we consider the following vector
\begin{equation}\label{vec3}
|\eta_{\theta}\rangle=\frac 1 4 \sum_{\ell=1}^4 |\xi_{\theta^{(\ell)}}\otimes \xi_{\theta^{(\ell)}}\rangle.
\end{equation}
For any fixed $i$ $(1\le i \le n)$, we choose $n$-tuple $\theta$ such that $\theta_i=0$ and $\theta_j=\pi/2$ unless $j=i$, then we see that the vector $|\eta_{\theta}\rangle$ in Eq.~\eqref{vec3} is identified with the matrix $|i\rangle\langle i|$.
 Now, for any fixed $i,\,j$ $(1\le i\neq j\le n)$, we choose $n$-tuple $\theta$ such that $\theta_i=\theta_j=0$ and $\theta_k=\pi/2$ unless $k=i$ and $k=j$. Then, the vector $|\eta_{\theta}\rangle$ in Eq.~\eqref{vec3} is idenitfied with the matrix $|i\rangle \langle i|+|i\rangle \langle j|+|j\rangle \langle i|+|j\rangle \langle j|$. Therefore, vectors $|\xi_{\theta}\otimes \xi_{\theta}\rangle$ in Eq.~\eqref{vec1} span all symmetric matrices $|i\rangle \langle i|$ and $|i\rangle \langle j|+|j\rangle \langle i|$ in $M_n$ under the identification between $\mathbb C^n\otimes \mathbb C^n$ and $M_n$. 

Finally, we observe that 
\[
\begin{aligned}
&|i\rangle \otimes |j\rangle \notin \mathcal V_i^{(n,k)}\, \text{ and }\ |j\rangle \otimes |i\rangle \notin \mathcal V_j^{(n,k)}\\
\Longleftrightarrow
&\, j=n-k+i\, \text{(mod }n)\, \text{ and }\, i=n-k+i\, \text{(mod }n)\\
\Longleftrightarrow
&\, 2k=0\, \text{ (mod }n)\\
\Longleftrightarrow
&\, k=n/2, 
\end{aligned}
\]
for $1\le k \le n-1$. Therefore, we can conclude that either $|i\rangle \otimes |j\rangle \in \mathcal V_i^{(n,k)}$ or $|j\rangle \otimes |i\rangle \in \mathcal V_j^{(n,k)}$ for each $1\le i \neq j\le n$, whenever $k\neq n/2$. 

By combining the above two results, we see that $W_{\Phi^{(n,k)}}^{\Gamma}$ has the spanning property. Consequently, we have the following theorem.
\begin{thm}For $n\ge 3$, $k=1,2,\ldots,n-1$, let $\Phi^{(n,k)}:M_n\to M_n$ be the positive linear map defined by Eq.~\eqref{map}. Then the entanglement witnesses $W_{\Phi^{(n,k)}}$ are optimal PPTES witnesses.  
\end{thm}

In conclusion, we have seen that paritial transposes of optimal entanglement witnesses $W_{\Phi^{(n,k)}}$ provided by X. Qi and J. Hou \cite{xia} are also optimal, and  $W_{\Phi^{(n,k)}}$ are  indeed optimal PPTES witnesses, that is, nd-OEW in the sense of \cite{lew00}. We note that indecomposable positive linear maps $\Phi^{(n,k)}$ are natural generalization of extremal Choi map between $M_3$. So  it would be meaningful to show whether these maps are extremal.

\begin{acknowledgments}
This work was partially supported by the Basic Science Research Program through the
National Research Foundation of Korea(NRF) funded by the Ministry of Education, Science
and Technology (Grant No. NRFK 2012-0002600)
\end{acknowledgments}


\end{document}